\documentclass[12pt]{article}
\usepackage[section]{placeins}

\begin{document}

\newcommand{\comment}[1]{}
\def\beq{\begin{equation}}
\def\eeq{\end{equation}}

\begin{center}

{\boldmath \bf $W$ boson polarization as a measure of gauge-Higgs anomalous
couplings at the LHC} 

\vskip .3cm

Kumar Rao \\
\vskip .1cm
{\it Physics Department, Indian Institute of Technology Bombay, \\Powai,
Mumbai 400076, India}\\
\vskip .1cm
Saurabh D. Rindani \\
\vskip .1cm
{\it Theoretical Physics Division, Physical Research Laboratory,
\\Navrangpura, Ahmedabad 380009, India} \\

\vskip 1cm
{\bf Abstract}

\end{center}
\vskip .3cm

{\footnotesize
We show how the $W$ boson polarization in the process of associated $W^{\pm}H$
production at the Large Hadron Collider (LHC) can be used to constrain
anomalous $WWH$ couplings. We first calculate the spin density matrix 
for the $W$ to linear order in the anomalous couplings, which are
assumed to be small. We then evaluate angular asymmetries in the decay
distributions of leptons produced in the decay of the $W$ and show how
they can be used to measure the individual elements of the polarization
tensor. We estimate the limits that can be placed on the anomalous $WWH$
couplings at a future run of the LHC. 
}

\vskip 1cm

\noindent {\bf 1. Introduction} \\
After the discovery of the Higgs boson with a mass of  around 125 GeV,
several measurements at the Large Hadron Collider (LHC) indicate that its couplings are
consistent with those predicted by the standard model (SM). However, a
complete confirmation that the Higgs boson $H$ discovered at the LHC is
indeed the Higgs boson of the SM will require precise determination of
all the couplings of $H$, including Higgs
self-couplings. A simplistic analysis, usually adopted in the
interpretation of Higgs data, attempts to measure the ratio $\kappa$ 
of the
coupling to that in the standard model. In this procedure, the so-called
$\kappa$ framework, the forms of the
interactions assumed are the same as in the SM at tree level. An attempt to
introduce more general tensor forms of couplings is not permitted by the
present accuracy of the experiments. However, in future experiments at
higher luminosities, it is hoped that such general forms of couplings
will be constrained. 
This could include measurement of differential cross sections, which
would
be highly data intensive. Alternatively, one could measure partial cross
sections, or angular or energy asymmetries of final state particles. 

An
interesting additional variable which we consider in this work is the 
polarization of the $W^{\pm}$ produced in association with the Higgs.  
Measurement of polarization of a heavy particle requires the observation
of decay distributions of the particle. Again one can construct
appropriate
asymmetries from the kinematical distributions of the decay particles.
In particular, charged lepton distributions in the decay of the $W$ would
enable the measurement of $W$ polarization parameters, which in turn
would constrain the strengths of the tensor structures of the $WWH$
interactions.

$W$ polarization has been discussed recently in the context of
polarized top decays and diboson resonances at the LHC \cite{aguilar},
and earlier in the context of various single, pair and associated $W$
production processes \cite{stirling}. For details of the formalism in
the context of LEP experiments, see \cite{bailey}.
$Z$ polarization has been studied in the context
of new physics at $e^+e^-$ colliders \cite{aguilar2,Rahaman:2016pqj}.

$W$ helicity fractions, which measure the degree of longitudinal or
transverse polarizations, have been measured in top decay $t\to bW$ 
at the LHC  from the polar-angle distributions, integrated
over the azimuthal angle \cite{helfrac}. These correspond to the diagonal elements of
the $W$ production spin-density matrix. In what follows, we also
consider measurement of the off-diagonal density-matrix elements
\cite{aguilar3,belyaev,Velusamy:2018ksp,Rahaman:2017qql,Boudjema:2009fz} through
angular asymmetries of the leptons produced in $W$ decay.

The asymmetries we consider are defined in the rest frame of the
decaying $W$. Measurement of these asymmetries would therefore involve
transforming laboratory-frame kinematic variables to the $W$ rest frame.
This in turn needs the knowledge of the $W$ four-momentum. This is a
potential problem because the $W$ decays into a neutrino, which is not
detected. While the transverse momentum of the neutrino can be
reconstructed with good accuracy using momentum conservation, the
longitudinal momentum cannot be measured directly. The usual procedure
\cite{helfrac} is to constrain the invariant mass of the $W$
decay products to be equal to the $W$ mass. Moreover, the construction 
of the polarization asymmetries, which are related to the elements of 
the $W$ density matrix requires the  $W$ to be on-shell 
\cite{Rahaman:2017qql,Boudjema:2009fz}.
Since the on-shell constraint gives rise to a quadratic equation, there
is a two-fold ambiguity in the determination of the neutrino
longitudinal momentum. Various procedures have been considered to choose
one of the two solutions allowed. One procedure followed in a recent
study of $WH$ production by ATLAS is to take the smaller of the two
solutions \cite{Aaboud:2017cxo}. Another suggestion 
\cite{Godbole:2014cfa} is to compare the longitudinal boosts $\beta_z^W$
and $\beta_z^H$ of the reconstructed $W$ and the $H$, and choose the
solution which gives the lower value for $\vert \beta_z^W - \beta_z^H
\vert $,
which was found in simulations 
to give the true neutrino momentum in 65\% of the cases.

$W$ and $Z$ polarization in associated Higgs production has been studied
recently in \cite{Nakamura:2017ihk}, with which our work has
considerable overlap. While \cite{Nakamura:2017ihk} contains expressions
for $W$ spin density matrices which we obtained independently, their
analysis deals with hadronic decay of the vector bosons, whereas we
concentrate on leptonic decay of the $W$. While the hadronic branching
ratios are larger, it is not possible to determine the charge of the
jets. On the other hand, though the branching ratio of $W$ into leptons
is smaller, greater precision is possible, as well as charge
discrimination is available.

The $WWH$ vertex for a process $W^{+*} \to W^+H$ may be written in a
model-independent way as
\beq\label{WWH}
\Gamma_{\mu\nu} = g m_W\left[a_W g_{\mu\nu} + \frac{b_W}{m_W^2} (q_\mu
k_\nu -
g_{\mu\nu} q\cdot k ) + \frac{\tilde b_W}{m_W^2}
\epsilon_{\mu\nu\alpha\beta}q^\alpha k^\beta \right],
\eeq
where $q$ is the incoming $W^*$ momentum and $k$ is the outgoing $W$
momentum, and $\nu$, $\mu$ are their respective polarization indices.
$g$ is the weak coupling constant, and $a_W =1$ in the SM at tree
level. $b_W$ and
$\tilde b_W$ which are vanishing in the SM at tree level, are anomalous couplings, taken to be complex form factors.
An analogous vertex for the process $W^{-*} \to W^-H$ may also be
written.
While the first two terms would arise from terms in an effective
Lagrangian and are invariant under CP, the $\tilde b_W$ term
would correspond to a CP-violating term in the Lagrangian. 
The anomalous couplings could arise at one or more loops in the SM, 
or in extensions of the SM, with heavy particles (the top quark, $W$, 
$Z$ and $H$ in the SM, or other additional particles in SM extensions) occurring
in the loops, and coupling to the Higgs boson. However, we will not be
concerned here with predictions of any specific model.
 
\bigskip
\noindent{\bf 2. Helicity amplitudes and density matrix} \\
We consider the process $pp \to W^{\pm}H X$ at the LHC, which at the
partonic level proceeds via the process $q\bar q' \to W^* \to W^{\pm}H$, where
$q$ and $q'$ are quarks. After calculating the helicity amplitudes for
the process in the presence of anomalous $WWH$ couplings, we evaluate
the production density matrix elements for the spin of the $W$ at the
partonic level and consequently for a hadronic initial state, to linear
order in the anomalous couplings. 
We further examine how each of these polarization tensor elements may be
measured from various angular asymmetries of charged leptons produced in
the decay of the $W$, and also estimate the sensitivity of these
measurements for an assumed integrated luminosity of the experiment.

To calculate the helicity amplitudes for the production process
in the quark-antiquark c.m.  (centre-of-mass) frame,
\beq\label{prodwplus}
u(p_1)+\bar d(p_2) \to W^+(k) + H,
\eeq
where $u$ and $d$ are respectively up-type and down-type quarks of any 
generation, 
we make use of the following representation for the polarization vectors
of the $W$:
\beq\label{epspm}
\epsilon^\mu(k,\pm) \equiv
\left(
0,\mp\frac{\cos\theta}{\sqrt{2}},-\frac{i}{\sqrt{2}},\pm\frac{\sin\theta}{\sqrt{2}}\right),
\eeq
\beq\label{eps0}
\epsilon^\mu(k,0) \equiv \left(\frac{|\vec k|}{m_W},\frac{E_W\sin\theta}{m_W},
	0,\frac{E_W\cos\theta}{m_W}\right)
\eeq
where $E_W$ is the energy of the $W$ and $\vec k_W$ its momentum, with
polar angle $\theta$ with respect to  the direction of the $u$ quark
taken as the $z$ axis. 

The nonzero helicity amplitudes in the limit of massless quarks are
given by
\beq\label{helamps1}
M(-,+,-) = -g^2V_{qq'}m_W\frac{\sqrt{\hat s}}{2}\left[a_W-  (b_W +i \beta_W \tilde
b_W)\frac{\sqrt{\hat s}E_W}{m_W^2}\right]\frac{(1+\cos\theta)}{(\hat s-m_W^2)}  
\eeq
\beq\label{helamps2}
M(-,+,0) = -g^2V_{qq'}\sqrt{\frac{\hat s}{2}}E_W\left[a_W -  b_W 
\frac{\sqrt{\hat s}}{E_W}\right]\frac{\sin\theta}{(\hat s-m_W^2)}
\eeq
\beq\label{helamps3}
M(-,+,+) = -g^2V_{qq'}m_W\frac{\sqrt{\hat s}}{2}\left[a_W -  (b_W -i \beta_W \tilde
b_W)\frac{\sqrt{\hat s}E_W}{m_W^2}\right]\frac{(1-\cos\theta)}{(\hat s-m_W^2)}
\eeq
where $\sqrt{\hat s}$ is the total energy in the parton c.m. frame, 
$\beta_W=|\vec k_W|/E_W$, and $V_{qq'}$ is the appropriate element
of the Cabibbo-Kobayashi-Maskawa matrix, and the first two entries in
$M$ correspond to helicities $-1/2$ and $+1/2$ of the quark and
anti-quark, respectively, and the third entry is the $W$ helicity.

The helicity amplitudes for the $W^-$ production process 
\beq\label{prodwminus}
d(p_1)+\bar u(p_2) \to W^-(k) + H
\eeq
are also given by eqns. (\ref{helamps1})-(\ref{helamps3}), with the first two entries in $M$
denoting the helicities of the $d$ and $\bar u$, and $\theta$
representing the angle between $W^-$ and $d$. Here it is assumed that
the same couplings $a_W$, $b_W$ and $\tilde b_W$ occur in the process
$W^{-*} \to W^-H$ as in $W^{+*} \to W^+H$, as in an effective field
theory approach \cite{Nakamura:2017ihk}.

In terms of the helicity amplitudes, the spin-density matrix for $W$
production is defined as
\beq\label{defrho}
\rho(i,j) = \overline{\sum_{h_q,h_{\bar q}}} M(h_q,h_{\bar q},i)M(h_q,h_{\bar q},j)^*,
\eeq
the sum and average being over initial helicities $h_q$, $h_{\bar q}$ of
the quark and anti-quark, respectively, and also over initial colour
states, not shown explicitly. 
The diagonal elements for $i=j$ would correspond to production
probabilities with definite $W$ polarization labelled by $i=j$ as
applicable, for example, in the study of helicity fractions. However,
in the description of $W$ production followed by decay, where
measurement is made on the decay products, the full density matrix
description, which includes off-diagonal elements, is needed.
This is because a full description requires multiplying the helicity
amplitudes for production with the helicity amplitudes for decay in a
coherent fashion (see, for example, \cite{Leader:2001gr}). 

The density matrix elements derived from the helicity amplitudes
(\ref{helamps1})-(\ref{helamps3}),
to linear order in the couplings $b_W$ and $\tilde b_W$, setting $a_W=1$
are as follows.
\beq\label{rhopmpm}
\rho(\pm,\pm) = \frac{g^4}{12}\frac{m_W^2 \hat s}{4(\hat s  - m_W^2)^2}|V_{qq'}|^2 
(1 \mp \cos\theta)^2 \left[ 1 - 2 ({\rm Re} b_W - \beta_W {\rm Im} \tilde
b_W ) \frac{\sqrt{\hat s}E_W}{m_W^2} 
\right]
\eeq
\beq\label{rho00}
\rho(0,0) = \frac{g^4}{12}\frac{E_W^2 \hat s}{2(\hat s  -
m_W^2)^2}|V_{qq'}|^2
\sin^2\theta \left[ 1 - 2 {\rm Re} b_W 
 \frac{\sqrt{\hat s}}{E_W}
\right]
\eeq
\begin{eqnarray}\label{rhomp0}
\rho(\mp,0)\!\! &\!\!=\!\!& \frac{g^4}{12}\frac{\hat s m_W
E_W}{2\sqrt{2}(\hat s - m_W^2)^2} |V_{qq'}|^2\sin\theta(1\pm \cos\theta)
\\ 
 &\hskip -1.7cm \times &  \hskip -1cm \left[1 
-{\rm Re}b_W \sqrt{\hat s}\frac{(E_W^2+m_W^2)}{E_Wm_W^2}
-i{\rm Im}b_W \sqrt{\hat s}
\frac{\beta_W ^2 E_W}{m_W^2}
\mp i\beta_W \tilde b_W \frac{\sqrt{\hat s}E_W}{m_W^2}
\right]
\end{eqnarray}
\beq\label{rhomppm}
\rho(\mp,\pm) = \frac{g^4}{12}\frac{m_W^2 \hat s}{4(\hat s  -
m_W^2)^2}|V_{qq'}|^2
\sin^2\theta \left[ 1 -  2({\rm Re} b_W \pm i \beta_W {\rm Re}
\tilde
b_W ) \frac{\sqrt{\hat s}E_W}{m_W^2}
\right]
\eeq

We have used the analytical manipulation software FORM \cite{form} to check these expressions.

Defining an integral of this density matrix over an appropriate kinematic
range as $\sigma(i,j)$, the latter can be parametrized in terms of the linear 
polarization $\vec P$ and the tensor polarization $T$ as follows.
\cite{Leader:2001gr}
\beq\label{vectensorpol}
\sigma(i,j) \equiv \sigma \;\left( 
\begin{array}{ccc}
\frac{1}{3} + \frac{P_z}{2} + \frac{T_{zz}}{\sqrt{6}} &
\frac{P_x - i P_y}{2\sqrt{2}} + \frac{T_{xz}-i T_{yz}}{\sqrt{3}} &
\frac{T_{xx}-T_{yy}-2iT_{xy}}{\sqrt{6}} \\
\frac{P_x + i P_y}{2\sqrt{2}} + \frac{T_{xz}+i T_{yz}}{\sqrt{3}} &
\frac{1}{3} - \frac{T_{zz}}{\sqrt{6}} &
\frac{P_x - i P_y}{2\sqrt{2}} - \frac{T_{xz}-i T_{yz}}{\sqrt{3}} \\
\frac{T_{xx}-T_{yy}+2iT_{xy}}{\sqrt{6}} &
\frac{P_x + i P_y}{2\sqrt{2}} - \frac{T_{xz}+i T_{yz}}{\sqrt{3}} &
\frac{1}{3} - \frac{P_z}{2} + \frac{T_{zz}}{\sqrt{6}} 
\end{array}
\right)
\eeq
where $\sigma(i,j)$ is the integral of $\rho(i,j)$, and $\sigma$ is the 
production cross section, 
\beq
\sigma = \sigma(+,+) + \sigma(-,-) + \sigma(0,0).
\eeq

The vector and tensor polarizations then can be obtained by inverting eqn. (\ref{vectensorpol}):
\begin{eqnarray}
P_x& =& \frac{1}{(\sqrt{2}\sigma)}[\sigma(+,0)+ \sigma(0,+)+\sigma(-,0)+\sigma(0,-)]\\
P_y& =& \frac{i}{(\sqrt{2}\sigma)}[\sigma(+,0)- \sigma(0,+)-\sigma(-,0)+\sigma(0,-)]\\
P_z& =& \frac{1}{\sigma}[\sigma(+,+)- \sigma(-,-)]\\
T_{xy}& =& \frac{i\sqrt{6}}{(4\sigma)}[\sigma(+,-)- \sigma(-,+)]\\
T_{xz}& =& \frac{\sqrt{3}}{(4\sigma)}[\sigma(+,0)+ \sigma(0,+)-\sigma(-,0)-\sigma(0,-)]\\
T_{yz}& =& \frac{i\sqrt{3}}{(4\sigma)}[\sigma(+,0)- \sigma(0,+)+\sigma(-,0)-\sigma(0,-)]\\
T_{xx}-T_{yy}& =& \frac{\sqrt{6}}{(2\sigma)}[\sigma(+,-)+ \sigma(-,+)]\\
T_{zz}& =& \frac{\sqrt{6}}{(6\sigma)}[\sigma(+,+)+ \sigma(-,-)-2 \sigma(0,0)],
\end{eqnarray}

\bigskip
\noindent {\bf 3. Leptonic asymmetries}\\
Obtaining spin information of the $W$ requires measurements to be made on
the decay products of the $W$. Using leptonic decays is more convenient
than using hadronic decays because charge identification is difficult,
if not impossible, for that latter case. Expressions may be obtained for
the decay-lepton distribution in the $W$ production process by combining
the relevant production-level density matrix elements with appropriate
decay density matrix elements and integrating over the appropriate phase
space. As mentioned before, a full measurement of the lepton distribution
would require a very large number of events. It is more economical to
use integrated angular asymmetries, which utilize all relevant events.
We therefore adopt this approach and define different angular
asymmetries of the charged lepton.

Following \cite{Rahaman:2016pqj}, we define angular asymmetries of the lepton
arising from $W$ decay, evaluated in the rest frame of the $W$,
 which isolate various elements of the
polarization tensor:
\begin{eqnarray}
A_x &=& \frac
{\sigma(\cos\phi^* > 0) - \sigma(\cos\phi^*<0)} 
{\sigma(\cos\phi^* > 0) + \sigma(\cos\phi^*<0)}, \\
A_y &=& \frac
{\sigma(\sin\phi^* > 0) - \sigma(\sin\phi^*<0)} 
{\sigma(\sin\phi^* > 0) + \sigma(\sin\phi^*<0)}, \\
A_z &=& \frac
{\sigma(\cos\theta^* > 0) - \sigma(\cos\theta^*<0)} 
{\sigma(\cos\theta^* > 0) + \sigma(\cos\theta^*<0)}, \\
A_{xy} &=& \frac
{\sigma(\sin2\phi^* > 0) - \sigma(\sin2\phi^*<0)} 
{\sigma(\sin2\phi^* > 0) + \sigma(\sin2\phi^*<0)}, \\
A_{xz} &=& \frac
{\sigma(\cos\theta^*\cos\phi^* < 0) - \sigma(\cos\theta^*\cos\phi^*>0)} 
{\sigma(\cos\theta^*\cos\phi^* > 0) + \sigma(\cos\theta^*\cos\phi^*<0)}, \\
A_{yz} &=& \frac
{\sigma(\cos\theta^*\sin\phi^* > 0) - \sigma(\cos\theta^*\sin\phi^*<0)} 
{\sigma(\cos\theta^*\sin\phi^* > 0) + \sigma(\cos\theta^*\sin\phi^*<0)}, \\
A_{x^2-y^2} &=& \frac
{\sigma(\cos2\phi^* > 0) - \sigma(\cos2\phi^*<0)} 
{\sigma(\cos2\phi^* > 0) + \sigma(\cos2\phi^*<0)}, \\
A_{zz} &=& \frac
{\sigma(\sin3\theta^* > 0) - \sigma(\sin3\theta^*<0)} 
{\sigma(\sin3\theta^* > 0) + \sigma(\sin3\theta^*<0)}. 
\end{eqnarray}
The direction of the quark momentum is defined as the $z$
axis, and the $x$ axis chosen so that the $W$ lies in the $xz$ plane.
Using these axes, the angles $\theta^*$ and $\phi^*$ are the polar and azimuthal angles of the
decay lepton, defined in the rest frame of the $W$, with respect to the
boost direction of the $W$.

It may be observed that since the sign of the triple vector product of
the beam direction, the $W$ momentum direction and the lepton momentum
direction determines the sign of  $\sin\phi^*$, the asymmetries $A_y$,
$A_{xy}$, $A_{yz}$ which are linear in $\sin\phi^*$ are measures of this
triple vector product. These asymmetries are therefore odd under naive
time reversal operation T$_{\rm N}$, which is simply reversal of all momentum
and spin directions. Hence these asymmetries would be either
proportional to the T-odd parameter $\tilde b_W$, or proportional to
the T-even coupling $b_W$, but to satisfy unitarity and the CPT
theorem, proportional only to its imaginary part. This will be seen in
the numerical expressions or asymmetries which follow later on.

The above results assume that the quark and antiquark directions can be
identified unambiguously. This is not true in the case of the LHC, where
the quark could arise from either proton, and the choice of the $z$ axis
is not unique. Taking into account the two possibilities when the quark
(and antiquark) arise from the two oppositely directed proton beams, we
find that the density matrix elements $\sigma(\pm,0)$ and
$\sigma(0,\pm)$ vanish, as also the polarizations $P_x$, $P_y$,
$P_{xz}$, $P_{yz}$ and the corresponding asymmetries $A_x$, $A_y$,
$A_{xz}$, $A_{yz}$. 

In what follows we will take the $z$ axis to be defined by the direction
of the reconstructed momentum of the combination $WH$. In this
case, the density matrix elements, polarizations and asymmetries which
were vanishing when the $z$ was chosen to be the beam direction now turn out
to be nonzero.
\bigskip

\noindent{\bf 4. Numerical results}\\
To start with,
we have evaluated the production spin density matrix elements after
integrating over the parton distribution functions as well as the
final-state phase space. We do not restrict ourselves to any particular
decay mode of the Higgs, but assume that full identification is
possible. In practice, one would have to apply kinematic
cuts for lepton identification, elimination of backgrounds, etc., as also
take into account the Higgs detection efficiency, which will require a
more refined analysis. 

We use the MMHT2014  parton distributions \cite{MMHT} with
factorization scale chosen as the square root of the partonic c.m.
energy.
For the two cases of $W^+$ and $W^-$ production, though the partonic
level cross sections and density matrices have the same expressions, the
parton densities corresponding to the initial states are different. Hence
the numerical results are different.

As mentioned before, we choose as $z$ axis the direction of the combined
momenta of $W$ and $H$.

The results for the density matrices for
$W^+$ production and $W^-$ production are shown
respectively in Table \ref{sigmaijplus} and Table \ref{sigmaijminus}. 
\begin{table}[h]
\centering
\begin{tabular}{|c|c|c|c|c|c|}
\hline
 & SM & Re~$b_W$ & Im~$b_W$ & Re~$\tilde b_W$ & Im~$\tilde b_W$ \\
\hline
$\sigma(\pm,\pm)$ & 165.8 & $-1757$ & 0 & 0 & $\mp 1273$ 
\\
$\sigma(0,0)$ & 388.7 & $-1757$ & 0 & 0 & 0
\\
$\sigma(\pm,\mp)$ & 82.91 & $-878.6$ & 0 & $\pm i 636.8$ & 0
\\
$\sigma(\pm,0)$ & 95.96 & $-872.8$ & $-i431.7$ & $\pm i 518.9$ & $\mp
518.9$
\\
$\sigma(0,\pm)$ & 95.96 & $-872.8$ & $i431.7$ & $\mp i 518.9$ & $\mp
518.9$
\\
\hline
\end{tabular}
\caption{\label{sigmaijplus}Production spin density matrix elements for the $W^+$ (in units
of fb) for the SM and the coefficients of various couplings in each
matrix element}
\end{table}

\begin{table}[h]
\centering
\begin{tabular}{|c|c|c|c|c|c|}
\hline
 & SM & Re~$b_W$ & Im~$b_W$ & Re~$\tilde b_W$ & Im~$\tilde b_W$ \\
\hline
$\sigma(\pm,\pm)$ & 110.2 & $-1140$ & 0 & 0 & $\mp 817.1$ 
\\
$\sigma(0,0)$ & 251.5 & $-1140$& 0 & 0 & 0
\\
$\sigma(\pm,\mp)$ & 55.10 & $-570.0$ & 0 & $\pm i 408.5$ & 0
\\
$\sigma(\pm,0)$ & 49.86 & $-439.6$ & $-i209.9$ & $\pm i 255.0$ & $\mp 255.0$
\\
$\sigma(0,\pm)$ & 49.86 & $-439.6$ & $i209.9$ & $\mp i 255.0$ & $\mp 255.0$
\\
\hline
\end{tabular}
\caption{\label{sigmaijminus}Production spin density matrix elements for the $W^-$ (in units
of fb) for the SM and the coefficients of various couplings in each
matrix element}
\end{table}

The total cross section for $W^+$ production has the expression
\beq\label{numcsplus}
\sigma = (720.2 - 5271\, {\rm Re}\, b_W)\, {\rm fb}.
\eeq
and that for $W^-$ production the expression
\beq\label{numcsminus}
\sigma = (471.8 - 3420\, {\rm Re}\, b_W)\, {\rm fb}.
\eeq
The total cross section for $W^+$ production could put a limit on
Re~$b_W$ of $2.28\times 10^{-4}$ with an integrated luminosity
$L=500\,{\rm fb}^{-1}$, and of $1.61\times 10^{-4}$ with $L=1000\,{\rm
fb}^{-1}$. The corresponding limits using cross section for $W^-$
production are $2.84\times 10^{-4}$ and $2.01\times 10^{-4}$.
Measurement of the cross section using only  electron and muon decay modes of the
$W^+$ assuming branching ratios of 10.71\% and 10.63\% respectively,
we can therefore set a limit of 
$4.93\times 10^{-4}$ on the coupling Re~$b_W$ for $L=500$~
fb$^{-1}$, and $3.49\times 10^{-4}$ for $L=1000$~ fb$^{-1}$. 
The corresponding numbers for $W^-$ are respectively $6.15\times
10^{-4}$ and $4.35\times 10^{-4}$.

The leptonic asymmetries corresponding to the different polarizations in
$W^+$ production and decay,
in an obvious notation, are given by
\beq A_x = -0.282 + 0.502\, {\rm Re}\,b_W \eeq 
\beq A_y = 1.52 \,{\rm Re}\,\tilde b_W \eeq 
\beq A_z = 2.60 \,{\rm Im}\,\tilde b_W \eeq 
\beq A_{xy} = -0.563\, {\rm Re}\,\tilde b_W \eeq
\beq A_{xz} = 0.649\,{\rm Im}\,\tilde b_W \eeq
\beq A_{yz} = 0.540\,{\rm Im}\,b_W \eeq
\beq A_{x^2 -y^2} = 0.0733 - 0.240  \,{\rm Re}\,b_W \eeq 
\beq A_{zz} = -0.116 - 0.849 \,{\rm Re}\,b_W \eeq
The corresponding asymmetries in $W^-$ production and decay are
\beq A_x = -0.224 + 0.351\, {\rm Re}\,b_W \eeq 
\beq A_y = 1.15 \,{\rm Re}\,\tilde b_W \eeq 
\beq A_z = 2.65 \,{\rm Im}\,\tilde b_W \eeq 
\beq A_{xy} = -0.551\, {\rm Re}\,\tilde b_W \eeq
\beq A_{xz} = 0.487\,{\rm Im}\,\tilde b_W \eeq
\beq A_{yz} = 0.401\,{\rm Im}\,b_W \eeq
\beq A_{x^2 -y^2} = 0.0744 - 0.230  \,{\rm Re}\,b_W \eeq 
\beq A_{zz} = -0.112 - 0.814 \,{\rm Re}\,b_W \eeq

As remarked earlier, the reconstruction of the $W$ rest frame in which
the above asymmetries are defined usually requires constraining the
$\ell\nu$
invariant mass to be equal to the $W$ mass. We have checked that if we do
not use this restriction and allow an off-shell $W$ to produce the
$\ell\nu$ pair, the asymmetries do not change by more than a few per
cent in most cases. Thus, the usual algorithms for constructing the $W$
rest frame would work with good accuracy.

In order to evaluate the 1-$\sigma$ limit $C_{\rm limit}$ on a coupling
$C$ which
can be obtained from the asymmetries, assuming one coupling to be
nonzero at a time, and an integrated luminosity $L$, we use the expression
\beq\label{sens}
C_{\rm limit} = \frac{\sqrt{1-A_{\rm SM}^2}}{\vert A - A_{\rm SM} \vert}
\frac{1}{\sqrt{\sigma_{\rm SM}L}},
\eeq
where $A$ is the asymmetry for unit value of the coupling $C$.
For $W^+$ production, 
for integrated luminosities of 500 fb$^{-1}$ and 1000 fb$^{-1}$, we obtain the
limits shown in Table \ref{limitswplus}.
\begin{table}[!htbp]
\centering
\begin{tabular}{|c|c|c|c|}
\hline
Asymmetry & Coupling & Limit (in $10^{-3}$)& Limit (in $10^{-3}$)\\
&& ($L=500~{\rm fb}^{-1}$) & ($L=1000~{\rm fb}^{-1}$)\\
\hline
$A_x$  &Re~$b_W$ &6.9&4.9
\\
$A_y$ &Re~$\tilde b_W$ &2.4&1.7
\\
$A_z$ &Im~$\tilde b_W$ &1.4&0.96
\\
$A_{xy}$ & Re~$\tilde b_W$ & 6.4&4.5
\\
$A_{xz}$ &Im~$\tilde b_W$  & 5.6&3.9
\\
$A_{yz}$ &Im~$b_W$ & 6.7&4.7
\\
$A_{x^2-y^2}$ &Re~$b_W$ & 15&11
\\
$A_{zz}$ &Re~$b_W$  & 4.2&3.0
\\
\hline
\end{tabular}
\caption{\label{limitswplus}1-$\sigma$ limits which could be obtained from
various leptonic asymmetries in $W^+$ production and decay, 
with integrated luminosities of 500 and 1000
fb$^{-1}$.}
\end{table}
The corresponding limits from $W^-$ production and decay are shown in
Table \ref{limitswminus}.
\begin{table}[!htbp]
\centering
\begin{tabular}{|c|c|c|c|}
\hline
Asymmetry & Coupling & Limit (in $10^{-3}$) & Limit (in $10^{-3}$)\\
&& ($L=500~{\rm fb}^{-1}$) & ($L=1000~{\rm fb}^{-1}$)\\
\hline
$A_x$  &Re~$b_W$ &12&8.7
\\
$A_y$ &Re~$\tilde b_W$ &3.9&2.7
\\
$A_z$ &Im~$\tilde b_W$ &1.7&1.2
\\
$A_{xy}$ & Re~$\tilde b_W$ & 8.1&5.7
\\
$A_{xz}$ &Im~$\tilde b_W$  & 9.2&6.5
\\
$A_{yz}$ &Im~$b_W$ & 11&7.9
\\
$A_{x^2-y^2}$ &Re~$b_W$ & 19&14
\\
$A_{zz}$ &Re~$b_W$  & 5.4&3.9
\\
\hline
\end{tabular}
\caption{\label{limitswminus}1-$\sigma$ limits which could be obtained from
various leptonic asymmetries in $W^-$ production and decay, 
with integrated luminosities of 500 and 1000
fb$^{-1}$.}
\end{table}

\FloatBarrier

The cross sections give the best limits on Re~$b_W$.
The results on the limits from leptonic asymmetries 
show that the asymmetries which are the most
sensitive ones are $A_{zz}$ for Re~$b_W$, $A_{yz}$ (the only one) 
for Im~$b_W$, $A_y$
for Re~$\tilde b_W$ and $A_z$ for Im~$\tilde b_W$. The limits from $W^+H$
production are better than those from $W^-H$ production in all cases.
However, it would be advantageous to combine results from both final
states to improve the results.

\bigskip
\noindent{\bf 5. Conclusions}\\
It is important to obtain
complete information about the Higgs boson discovered at the LHC,
including the tensor form of the couplings.
A proposal to measure form and magnitude of  
the coupling of the Higgs boson to a pair of $W$ bosons through the
polarization data of the $W$ is investigated here. The polarization
density matrix elements of the $W$ can be measured through certain
angular asymmetries of the charged lepton produced in $W$ decay, and we have
studied the sensitivity of these asymmetries to the anomalous couplings
$b_W$ and $\tilde b_W$ defined in eqn. (\ref{WWH}). Our results for $W^+$
and $W^-$ are shown
in tables \ref{limitswplus} and \ref{limitswminus}.

We see that a high degree of accuracy could be obtained in the
measurement of the $WWH$ anomalous couplings from the measurement of the
$W$ polarization parameters through suitable angular asymmetries of
leptons assuming an integrated luminosity of 500 fb$^{-1}$. There is
considerable improvement, as expected, if the luminosity is increased to
1000 fb$^{-1}$. The 1-$\sigma$ limits in most cases are of the order of a
few times $10^{-3}$. 

As mentioned earlier, the angular asymmetries we discuss are defined in
the rest frame of the $W$. The reconstruction of the $W$ rest frame in
the presence of the undetected neutrino has its drawbacks, and would
entail some loss in efficiency. We have also not taken into account
acceptance and isolation cuts on leptons. We also assume 100\% efficiency
for the detection of the Higgs. 
To get some idea of the effect of cuts,
we did evaluate the angular asymmetries and the sensitivities in the
presence of generic LHC acceptance cuts on the transverse momentum and 
the rapidity of the leptons. We found that the asymmetries do not change
much.
A full-scale
analysis using an event generator coupled with all appropriate cuts relevant
to the decay channels of the Higgs would be able to refine the actual
sensitivities that we have obtained.  It would also be profitable to
combine the results from $W^+$ and $W^-$ production processes, which
would improve the accuracy.

\vskip .2cm
\noindent {\bf Acknowledgement}: We thank Pankaj Sharma for collaboration in the initial stages of the work. KR acknowledges support from IIT Bombay, grant no. 12 IRCCSG032. SDR acknowledges support from the Department of
Science and Technology, India, under the J.C. Bose National
Fellowship programme, Grant No. SR/SB/JCB-42/2009. We thank Rohini
Godbole for discussions. We thank the referee for improvements and 
the suggestion for the choice of $z$ axis. 

\end{document}